\documentclass[12pt,preprint]{aastex}
\usepackage{graphicx}
\usepackage{emulateapj5}           
\slugcomment{Accepted for the Astrophysical Journal, likely ApJ 10 October 2005, v632 1 issue}

\shorttitle{SMA submm line observations of Orion-KL}
\shortauthors{Beuther et al.}

\begin{document}


\title{Submm line imaging of Orion-KL with the Submillimeter Array}


\author{H.~Beuther, Q.~Zhang, L.J.~Greenhill, M.J.~Reid, D.~Wilner, E.~Keto, , H.~Shinnaga, P.T.P.~Ho, J.M.~Moran,}

\affil{Harvard-Smithsonian Center for Astrophysics, 60 Garden Street, Cambridge, MA 02138, USA}

\email{hbeuther@cfa.harvard.edu}

\author{S.-Y.~Liu, C.-M.~Chang}
\affil{Academia Sinica Institute of Astronomy and Astrophysics,  No.1, Roosevelt Rd, Sec. 4, Taipei 106, Taiwan, R.O.C.}







\begin{abstract} 
We present the first submm (865\,$\mu$m) imaging spectral line survey
at one arcsecond resolution conducted with the Submillimeter Array
toward Orion-KL. Within the $\rm{two}\,\times\,\rm{two}$\,GHz
bandpasses (lower and upper sidebands, 337.2--339.2\,GHz and
347.2--349.2\,GHz), we find about 145 spectral lines from 13 species,
6 isotopologues, and 5 vibrational excited states.  Most
nitrogen-bearing molecules are strong toward the hot core, whereas the
oxygen-bearing molecules peak toward the south-west in the so-called
compact ridge. Imaging of spectral lines is shown to be an additional
tool to improve the identifications of molecular lines. Arcsecond
spatial resolution allows us to distinguish the molecular line
emission of the sources {\it I} and {\it n} from that of the hot
core. The only molecular species detected strongly toward source {\it
I} is SiO, delineating mainly the collimated north-east south-west
low-velocity outflow. The two positions close to source {\it I}, which
have previously been reported to show maser emission in the v=0
$^{28}$SiO(1--0) and (2--1) lines, show no detectable maser emission
in the v=0 $^{28}$SiO(8--7) line at our spatial resolution. SiO is
weak toward source {\it n}, and thus source {\it n} may not currently
be driving a molecular outflow. CH$_3$OH is the molecule with the
highest number of identified lines (46) in this spectral window. This
``line forest'' allows us to estimate temperatures in the region, and
we find temperatures between 50 and 350\,K, with the peak temperatures
occurring toward the hot core. The detection of strong vibrational
excited line emission from the submm continuum peak SMA1 supports the
interpretation that the source SMA1 is likely of protostellar nature.
\end{abstract}


\section{Introduction}

Orion-KL is the most studied region of massive star formation. At a
distance of $\sim$450\,pc, its molecular spectral line emission is
very strong (e.g., \citealt{schilke1997b}). Most molecular line
studies toward Orion-KL have been carried out with single-dish
instruments (e.g.,
\citealt{sutton1985,blake1987,schilke1997b,schilke2001,comito2005})
and thus do not resolve the spatial distribution of the molecular
gas. Molecular line studies with interferometers have been conducted
only at wavelengths longer than 1\,mm (e.g.,
\citealt{wilner1994,wright1996,blake1996,wilson2000,liu2002}). While 
many such studies were single-molecule observations, the first
dedicated interferometric line-surveys were conducted by
\citet{wright1996} and \citet{blake1996}. The interferometric
investigations found that the single-dish spectral features known as
the {\it hot core}, {\it compact ridge} and {\it plateau} are all
associated with the KL region whereas the dust continuum source CS1
about $25''$ north-east of source {\it I} exhibits narrower line
widths, lower temperatures and only weak indications of star formation
(e.g., \citealt{wright1996}).

The hot core as traced by dust continuum emission and nitrogen-bearing
molecules such as CH$_3$CN or NH$_3$ appears as a chain of dense
clumps offset by $\geq 1''$ from the radio source {\it I} with
estimated temperatures between 130 and 335\,K (e.g.,
\citealt{wilner1994,wright1996,wilson2000}).
The compact ridge a few arcseconds south-west of source {\it I} is
reported to have lower temperatures (of the order 100\,K) and to be
especially prominent in oxygen-bearing molecules (e.g.,
\citealt{wright1996,liu2002}). Furthermore, the region exhibits a
complex cluster of infrared sources studied from near- to mid-infrared
wavelengths \citep{dougados1993,greenhill2004,shuping2004}. At least
two outflows are driven from the region on scales $>10^4$\,AU, one
high-velocity outflow in the south-east north-west direction observed
in molecular lines and in the optical and near-infrared (e.g.,
\citealt{allen1993,wright1995,chernin1996,schultz1999}), and one lower
velocity outflow in the north-east south-west direction best traced by
thermal SiO and H$_2$O maser emission as well as some H$_2$ bow shocks
(e.g.,
\citealt{genzel1989,blake1996,chrysostomou1997,stolovy1998}). The
driving source(s) of the outflows are uncertain: initial claims that
it might be IRc2 are outdated now, and possible culprits are the radio
sources {\it I} and/or the infrared source {\it n}, also known as
radio source {\it L} \citep{menten1995}. Radio source {\it I} lies
close to the center of the north-east south-west outflow
\citep{gezari1992,menten1995,gezari1998,greenhill2003} and has not
been detected in the near- to mid-infrared \citep{greenhill2004}.

A study of the spatial distribution of the different molecular lines
can be used to identify and separate different centers of activity via
their different excitation conditions. The Submillimeter Array
(SMA\footnote{The Submillimeter Array is a joint project between the
Smithsonian Astrophysical Observatory and the Academia Sinica
Institute of Astronomy and Astrophysics, and is funded by the
Smithsonian Institution and the Academia Sinica.}) on Mauna Kea was
dedicated in November 2003, and Orion was one of its early science
targets. Therefore, we present the first high-spatial-resolution
($\sim 450$\,AU) submm wavelength molecular line investigation at
865\,$\mu$m. Since the dataset is incredibly rich, a complete analysis
and interpretation of all existing features is out of the scope of
this paper. Here, we present the overall content of the dataset and
focus on a few aspects in more detail. Follow-up investigations are
currently undertaken, and the data are publicly available (raw data
via the SMA web-site http://cfa-www.harvard.edu/rtdc/index-sma.html,
and calibrated data from the lead author of this paper).

The submm continuum emission was presented in a separate paper
\citep{beuther2004g}. The continuum data resolved source {\it I} from
the hot core and detected, for the first time, source {\it n} at a
wavelength shorter than 7\,mm. Furthermore, a new continuum peak
between the sources {\it I} and {\it n} is found, SMA1, which is
either an independent deeply embedded protostellar object or part of
the more extended hot core. These submm continuum sources, reproduced
in Figure \ref{continuum}, will be used as a reference frame for the
discussion of the line emission throughout this paper.

\section{Observations}
\label{obs}

Orion-KL was observed with the Submillimeter Array (SMA) in December
2003 and February 2004 at 348\,GHz ($865\,\mu$m) with 7 antennas in
two configurations and with projected baselines between 15 and
230\,k$\lambda$. The phase center was the nominal position of source
{\it I} as given by \citet{plambeck1995}: R.A. [J2000]
$5^h35^m14.50^s$ and Dec. [J2000] $-5^{\circ}22'30''.45$. For bandpass
calibration we used Jupiter and Mars. The flux scale was derived from
observations of Callisto and is estimated to be accurate to within
15\%. Phase and amplitude calibration was done via frequent
observations of the quasar 0420-014 about 17$^{\circ}$ from the phase
center. The zenith opacities, measured with the NRAO tipping
radiometer located at the Caltech Submillimeter Observatory, were
excellent during both tracks with $\tau(\rm{348GHz})\sim 0.1$ \& 0.14
(scaled from the 225\,GHz measurement via $\tau(\rm{348GHz})\sim
2.8\times \tau(\rm{225GHz})$). The receiver operated in a
double-sideband mode with an IF band of 4-6\,GHz so that the upper and
lower sideband were separated by 10\,GHz. The correlator had a
bandwidth of 2\,GHz and the spectral resolution was
0.8125\,MHz. Measured double-sideband system temperatures corrected to
the top of the atmosphere were between 250 and 600\,K, mainly
depending on the elevation of the source.

The uv-coverage with no projected baselines below 15\,k$\lambda$
implies that source structure on spatial scales $\geq 13''$ is
filtered out by the observations. Interferometric observations of
sources close to Declination $0^{\circ}$ like Orion-KL result in a
rather poor uv-coverage and dirty beam (Fig.~\ref{dirty}), further
complicating the image deconvolution. Since there is significant
large-scale emission associated with most spectral lines, we
deconvolved the Point Spread Function with the CLEAN-algorithm over a
large spatial area ($26''\times 26''$, about 2/3 of the primary beam
size) for each spectral line to derive the final images. It is
possible that, for some spectral lines, specific box-cleaning could
improve the final images but this is beyond the scope of this
paper. We applied different uv-weightings for various lines depending
mainly on the line strength, and obtained average synthesized beams of
$1.1''\times 0.9''$ (P.A. $-44^{\circ}$).

To get a feeling for the amount of missing flux, we convolved the data
to a $20''$ beam and compared them with the single-dish line survey by
\citet{schilke1997b}. Many strong and extended lines have missing flux 
values up to 90\%. For one of our main target lines, $^{28}$SiO(8--7),
the measured SMA flux is even lower, $\leq 10\%$ of that measured with
the CSO (see also the discussion at the end of \S \ref{sio}). Since
this missing flux has to be distributed over scales $>13''$, possibly
as large as the $20''$ beam of the CSO, the missing $^{28}$SiO(8--7)
flux scaled to the $1''$ synthesized SMA beam can be estimated to
approximately 3\,Jy\,beam$^{-1}$. For additional discussion on missing
flux in particular lines see \S \ref{sio} \& \ref{ch3oh}.

The sensitivity was dynamic-range limited by the side-lobes of the
strongest emission peaks and thus varied between the line maps of
different molecules and molecular transitions. This limitation was due
to the incomplete sampling of short uv-spacings and the presence of
extended structures. The theoretical $1\sigma$ rms per 1\,km\,s$^{-1}$
channel was $\sim 100$\,mJy whereas the measured $1\sigma$ rms in a
1\,km\,s$^{-1}$ channel image was of the order 500\,mJy because of the
dynamic-range problem and thus insufficient cleaning. The $1\sigma$
rms for the velocity-integrated molecular line maps (the velocity
ranges for the integrations were chosen for each line separately
depending on the line-widths and intensities) ranged between 70 and
320\,mJy with a mean $1\sigma$ value of $\sim 170$\,mJy. We calibrated
the data within the IDL superset MIR developed for the Owens Valley
Radio Observatory and adapted for the SMA; the imaging was performed
in MIRIAD. For more details on the array and its capabilities see
\citet{ho2004}.

\section{Results}
\label{results}

Figure \ref{spectra_all} presents the whole observed bandpass with its
extremely rich line forest, a closer zoom into each sideband is shown
in Figures \ref{spectra_lower} \& \ref{spectra_upper}. The line
identifications are not always unambiguous because different velocity
components and large line-widths are present in the region. At first,
the lines were identified by comparing our spectra with those of the
previous single-dish study of \citet{schilke1997b}. Then, in selected
cases we used the JPL, CDMS and LOVAS line catalogs
\citep{poynter1985,mueller2001,lovas2004} for confirmations and further
identifications. Altogether, we find about 145 lines above an
approximate flux level of 8\,Jy on the short baseline of 21\,m
(Figs.~\ref{spectra_all}, \ref{spectra_lower} \&
\ref{spectra_upper}). We note that this is not the sensitivity limit
of the data but rather a cutoff below which confusion becomes an even
larger problem and line identifications get considerably more
difficult. Of the 145 lines listed in Table \ref{lines}, we are able
to identify approximately 90\%. Tentative identifications are marked
in Table \ref{lines} and Figures \ref{spectra_all},
\ref{spectra_lower}, \ref{spectra_upper}, \& \ref{images} with ``?''. 

In addition to the listed lines, especially the upper sideband data
(Fig.~\ref{spectra_upper}) show many features at low intensities, of
which a majority is likely not noise but consists of weak additional
spectral lines. At the given noise level, we do not investigate these
features any further, but it is likely that finding real line-free
spectral regions to study the pure continuum emission in such hot core
type sources is difficult. The produced continuum emission will mostly
be ``polluted'' by underlying weak line emission. As outlined below
and in \citet{beuther2004g}, imaging of the lines helps significantly,
e.g., source {\it I} is nearly line-free (except for SiO) and thus the
continuum emission is reliable, whereas this is less likely the case
for the hot core region. To get an estimate of the line contamination,
we produced a second pseudo-continuum image averaging the whole upper
sideband data (in contrast to the continuum image presented in
\citet{beuther2004g} and Fig.~\ref{continuum}, where we excluded all
strong lines from the upper sideband). We did not use the lower
sideband data because the line contamination there is significantly
higher due to the strong CH$_3$OH bands. Deriving the source peak flux
values from this new image for each continuum sub-source, the values
are only 5-10\% higher than for the original image. While the original
image still covers a considerable bandpass of the upper sideband with
low-level emission, it is unlikely that the low-level emission exceeds
the contribution of the strong lines in the bandpass. Therefore, we
estimate the continuum flux uncertainties due to line contamination in
the image presented in
\citet{beuther2004g} and Fig.~\ref{continuum} to be $<10\%$, 
independent of the sub-sources. This is well within our calibration
uncertainties of 15\% (\S \ref{obs}). We note that this line
contamination estimate is only valid for this particular dataset and
cannot be extrapolated to single-dish studies, because the spatial
filtering for the line and continuum data with the interferometer
varies and single-dish bolometers cover a significantly larger
bandwidth (e.g., $\sim 80$\,GHz for the MAMBO array,
\citealt{kreysa1998})

Fifteen percent of the lines listed in Table \ref{lines} were not
reported in the previous single-dish survey \citep{schilke1997b}, not
to mention the additional low-level emission. In contrast, we do not
identify any clear interferometric non-detection of a previously
detected single-dish line, which could occur for very broadly
distributed molecular gas, implying that most molecular emission must
have some compact components.

Many of the lines listed in Table \ref{lines} can be attributed to a
few molecules, and we identify 13 different species, 6 isotopologues,
and 5 vibrational excited states within the dataset (Table
\ref{species}). Most of these have clearly separated lines without
much blending and thus can be imaged. Choosing one of the strongest
lines of each species, we produced velocity integrated images of
each. Figure
\ref{images} presents a compilation of line images from representative
species in the dataset. A first inspection of these line images
highlights the following interesting characteristics:

\begin{itemize}

\item
The main molecular species detected toward source {\it I} is SiO. It
clearly delineates the known low-velocity north-east south-west
outflow. Other species are either not detected toward source {\it I}
(e.g., CH$_3$OH) or exhibit only weak emission (e.g., SO$_2$).

\item
Most emission from nitrogen-bearing species is strong toward the hot
core and has a similar morphology to NH$_3$
(\citealt{wilson2000}). This confirms previous results at lower
frequencies by \citet{wright1996} and \citet{blake1996}.

\item 
Most oxygen-bearing molecules are weaker toward the hot core but
stronger in the south-west toward the so-called compact
ridge\footnote{The ``compact ridge'' has no clearly defined peak
positions but extends $5''-10''$ in east-west direction with peak
positions varying with the molecular line transitions (see
Fig.~\ref{images}).} (see also \citealt{wright1996}).

\item 
Sulphur-bearing species show emission toward the hot core and the
compact ridge$^2$ (see also \citealt{wright1996}).  Vibrationally
excited lines are stronger toward the hot core and SMA1.

\item 
Both, oxygen- and sulphur-bearing molecules show additional emission
toward the north-west, spatially associated with IRc6. Millimeter
continuum emission from this region has been previously reported by
\citet{blake1996}.

\item
The imaging of the molecular lines allows a better identification of
the spectral lines. For example, the previous identification of the
HCOOH$(15_{4,12}-14_{4,11})$ line at 338.144\,GHz \citep{schilke1997b}
is called into question because its spatial structure resembles rather
that of an N-bearing than of an O-bearing species. Its spatial
distribution appears also different to its 1\,mm counterpart presented
in \citet{liu2002}. Therefore, this line is more likely
CH$_3$CH$_2$CN$(37_{3,34}-36_{3,33})$. Some of the tentative
identifications in Table \ref{lines} and Figures \ref{spectra_all},
\ref{spectra_lower}, \ref{spectra_upper}, \& \ref{images} are uncertain 
due to this as well.

\end{itemize}

Another way to study the spatial differences of the molecular line
emission is via extracting spectra from the whole spectral data-cube
at selected positions within the field of view. For this purpose, we
imaged the large data-cube with a lower spectral resolution of
3\,km\,s $^{-1}$ and extracted spectra toward the positions of source
{\it I}, source {\it n}, the hot core, SMA1, three CH$_3$CN peak
positions, the southern CH$_3$OH peak position, the prominent western
gas peak and the most northern peak which is strong in HC$_3$N
(Fig.~\ref{spectra_positions}).

The most striking difference appears to be between source {\it I} and
most of the other positions. Toward source {\it I}, only $^{28}$SiO
and its rarer isotopologue $^{30}$SiO are strong, whereas toward the
other positions many more lines are detected but SiO is weaker. The
hot core and SMA1 as well as the CH$_3$CN peak positions are all
strong line emitters. Regarding CH$_3$OH, the strongest and most
prominent position is the southern peak associated with the compact
ridge. It shows the strongest CH$_3$OH line bands in the ground state
as well as the vibrationally excited states. The north-western peak is
still rather strong in the ground state CH$_3$OH lines but the
vibrationally excited bands are significantly weaker than toward the
southern peak.

A comparison of these different spectra shows the necessity of high
spatial resolution in such complex regions. At lower angular
resolution, the complexity of spatial variations and different
excitation conditions blends together and results in integrated spectra
as seen in many single-dish studies and also in
Fig.~\ref{spectra_all}.

\section{Analysis and Discussion}


\subsection{$^{28}$SiO(8--7) and $^{30}$SiO(8--7)}
\label{sio}

The main $^{28}$SiO(8--7) line was one of our prime target lines
within this survey. Previous studies of the lower excitation ground
state $^{28}$SiO lines showed that in Orion-KL SiO traces the
collimated north-east south-west outflow as well as the less
collimated structures in the north-west south-east direction
\citep{chandler1995,wright1995,blake1996}. A peculiarity of the
$^{28}$SiO v=0 emission is that the (1--0) and the (2--1) lines show
maser emission in a bow-tie-like structure close to source {\it I}
\citep{chandler1995,wright1995}. The orientation of the bow-tie is
along the axis of the north-east south-west outflow, its two peak
positions are approximately $0.5''$ offset from source {\it I} in both
directions. The $^{28}$SiO(5--4) observations of \citet{blake1996}
lacked the spatial resolution to disentangle the potential maser
positions from the other SiO emission, but at their given spatial
resolution they do not find extremely high brightness temperatures
which would imply obvious $^{28}$SiO(5--4) maser emission. We now
resolve the potential maser positions better in the $^{28}$SiO(8--7)
line but we also do not find any indication for $^{28}$SiO(8--7) maser
emission toward these positions. 

However, there is thermal $^{28}$SiO(8--7) and $^{30}$SiO(8--7)
emission on larger spatial scales tracing the molecular
outflows. Figures \ref{sio_channel} \& \ref{30sio_channel} show
channel maps of the $^{28}$SiO and $^{30}$SiO(8--7) data. While the
velocity range from $-5$ to 20\,km\,s$^{-1}$ is dominated by the
collimated north-east south-west outflow, the higher velocity
channels, especially of $^{28}$SiO(8--7), also show more extended
emission. We stress that the $^{28}$SiO(8--7) emission suffers
strongly from missing short spacings and thus side-lobe
problems. Especially, the north-east south-western stripes at the
edges of the field presented in the $^{28}$SiO(8--7) channel map
(Fig.~\ref{sio_channel}, the middle row) are side-lobe emission and not
real. Comparing the channel maps with a vector-averaged
$^{28}$SiO(8--7) spectrum (Fig.~\ref{sio_spectrum}), one finds that
the collimated and the extended emission features are clearly
separated in the spectrum as well. The more extended features are
represented by the secondary peaks at $<-5$ and $>20$\,km\,s$^{-1}$,
whereas the collimated north-east south-west structure is confined to
the lower velocities in between.

To investigate the different distributions, we averaged the
$^{28}$SiO(8--7) emission over the red and blue secondary peaks, the
red and blue collimated outflow, and the central velocities. The final
images are presented in Figure \ref{sio_sample}. We used the main
isotopologue for this purpose because, even considering the short
spacings problems, it is more sensitive to the larger-scale
emission. The collimated structures at low to intermediate velocities
are similar for both isotopologues. Obviously, the central
velocities are dominated by the collimated outflow
(Fig.~\ref{sio_sample}\,left). Going to slightly higher velocities we
find blue and red emission toward the south-west and the north-east of
source {\it I} (Fig.~\ref{sio_sample}\,middle). Since we resolve the
outflow spatially well in north-east south-west direction and still
find significant overlap of blue and red emission on both sides of
source {\it I}, this indicates that the outflow is likely close to the
plane of the sky (For an outflow orientation which is close to the
plane of the sky, expanding outflow lobes along the line of sight
produce the observed red and blue signature.). Going to the higher
velocity secondary peaks (Fig.~\ref{sio_sample}\,right), the emission
is more extended and spatially consistent with previous
$^{28}$SiO(2--1) observations by \citet{wright1995}. The blue emission
might be part of the larger-scale north-west south-east outflow but
the signature of the red higher velocity $^{28}$SiO emission is less
clear. It shows emission toward the south-east as expected from the
large-scale outflow, but it has additional emission toward the
south-west as well. As already mentioned, the $^{28}$SiO emission
suffers from missing flux problems due to the missing short spacings
(Figs.~\ref{sio_channel} \& \ref{sio_sample}), and it is difficult to
interpret the large-scale emission in more detail.

Source {\it n} may be the driver of one of the outflows in the region
(e.g., \citealt{menten1995}). However, source {\it n} is a strong
near-infrared source and thus not deeply embedded and not particularly
young. In addition, source {\it n} is weak in the usually
outflow-tracing SiO emission (Figs.~\ref{images} \&
\ref{spectra_positions}), which might be indicative of no recent 
molecular outflow activity. In contrast to this, there is indicative
evidence for a potential outflow from source {\it n} based on radio,
H$_2$O maser and infrared emission signatures (e.g.,
\citealt{menten1995,stolovy1998,greenhill2004,shuping2004}). While the 
detection of SiO emission often traces molecular outflows (e.g.,
\citealt{schilke1997a,gueth1999,cesaroni1999}), the SiO non-detection 
close to source {\it n} does not necessarily imply the opposite. For
example, dissociative shocks with velocities $>60$\,km\,s$^{-1}$ do
not produce significant SiO emission (e.g., \citealt{flower1996}), and
SiO can also be found in distinct bullets further down from the
outflow center. However, it remains interesting to note that the SiO
emission from the sources {\it I} and {\it n} is so different. Since
source {\it n} is less deeply embedded than source {\it I}, it may
either have driven an outflow in the past and we might still observe
the shocked remnants in H$_2$O maser emission \citep{menten1995}, or
the outflow is potentially of a slightly different, maybe more evolved
nature than the ones usually observed in SiO.

Figure \ref{sio_spectra} shows a comparison of the $^{28}$SiO and
$^{30}$SiO spectrum integrated over the central $4''\times 4''$
region. Both spectra cover about the same velocity range, and the
$^{30}$SiO spectrum is only $20\%$ weaker than the main isotopologue
$^{28}$SiO. This is different from the line ratio observed for these
lines with the CSO 10.4\,m single-dish telescope. \citet{schilke1997b}
find in their survey a $^{28}$SiO/$^{30}$SiO ratio of about 8,
consistent with a $^{28}$SiO opacity of 2. It appears that the
observed line ratio decreases at higher spatial resolution. However,
missing short spacings are a severe problem in comparing fluxes from
different species and isotopologues because the spatial filtering
affects various species differently. Trying to quantify this effect,
we convolved the $^{28}$SiO and $^{30}$SiO maps to the $20''$
resolution of the CSO observations and compared the resulting peak
fluxes. Additionally, we did the same again but this time masking out
all negative features. The first approach gives an extreme lower limit
to the recovered line intensities because the negative flux caused by
the missing large-scale emission is included (Table
\ref{sio_intensities}). With these two approaches, we recover between 
0.3 and 13\% of the $^{28}$SiO line flux and between 15 and 49\% of
the $^{30}$SiO line flux. An additional effect caused by the missing
short spacings is that the outflow emission is embedded in a
larger-scale ``bowl'' of negative emission which to first order lowers
the observed peak fluxes as well. Trying to estimate the distortion of
the measured fluxes by this ``bowl'', we find that the main $^{28}$SiO
line intensity is reduced by approximately $15\%$ whereas the rarer
$^{30}$SiO is affected by the ``bowl'' only to about $5\%$. While it
is difficult to give exact numbers how much emission is filtered out
in each line, these estimates show that the spatial filtering affects
the main isotopologue $^{28}$SiO significantly stronger than its rarer
species $^{30}$SiO. As outlined in \S \ref{obs}, the missing flux has
to be distributed over large spatial scales, and compared to the
approximate missing flux of 3\,Jy per synthesized SMA beam for
$^{28}$SiO (\S
\ref{obs}), this value is considerably lower for $^{30}$SiO, between
0.21 and 0.35\,Jy per synthesized SMA beam (depending on the SMA
$^{30}$SiO flux measurement in Table \ref{sio_intensities}). In
addition to this, it is likely that the line opacities toward the
central $(4'')^2$ are higher than in the larger region traced by the
CSO ($20''$ beam). This would also decrease the observed
$^{28}$SiO/$^{30}$SiO ratio. Hence, the observed $^{28}$SiO/$^{30}$SiO
ratio of approximately unity is explicable by a combination of spatial
filtering and opacity effects.

\subsection{CH$_3$OH emission}
\label{ch3oh}

The methanol molecule shows by far the most lines within our spectral
coverage. We identify 49 transitions from the $v_t =0,1,2$ states of
CH$_3$OH as well as its isotopologue $^{13}$CH$_3$OH (46 and 3,
respectively).  As shown in Figure \ref{images}, the methanol spatial
distribution is significantly different from N-bearing hot core
molecules with a strong additional peak in the south-west toward the
compact ridge. We also find that the CH$_3$OH emission is stronger
toward the SMA1 submm continuum peak than toward the strongest submm
continuum peak, which usually determines the hot core peak position.

The rich CH$_3$OH spectra not only provide spatial and kinematic
information but also allow possible estimates of the excitation
condition of the gas throughout the region. The observed CH$_3$OH
transitions have upper state energy levels spanning from $\approx$ 75K
to $\approx$ 700 K, readily indicating the high-temperature nature of
the gas in this region.  Assuming the CH$_3$OH line emission
originates from the same parcels of gas along each line of sight in
local thermodynamic equilibrium (LTE, thus having the same size,
velocity, and a single temperature), rotational temperatures can be
derived by the rotation or population diagram analysis (e.g.,
\citealt{goldsmith1999}). However, line blending among the CH$_3$OH
transitions as well as with other species complicates the line
intensity estimates and subsequent analysis. To remedy the situation,
we estimate the temperature by optimizing synthetic spectra to
observed spectra, an approach similar to those employed by, for
example, \citet{nummelin1998} and \citet{comito2005}.  First, the
observed CH$_3$OH lines toward a particular line-of-sight are
considered independent of their intensities, associated only by their
relative rest frequency differences. By using the minimum $\chi^2$
algorithm between the synthetic and observed spectra, we fitted all
CH$_3$OH lines simultaneously with Gaussian line profiles to derive
common center velocities and linewidths.  With the fixed velocities
and linewidths, the rotational temperatures are then derived via a
second $\chi^2$ minimization, matching the synthetic to the observed
intensities. In this derivation, the line opacity is assumed to have
Gaussian profiles, a beam filling factor is employed, and the
background radiation has been ignored.

Figure \ref{sample} presents observed and synthetic (fitted) spectra
toward the hot core position and a location in the vicinity of the
compact ridge but offset from the torsionally excited lines (see also
Fig.~\ref{trot}, the position is approximately the center of the HCOOH
emission presented in \citealt{liu2002}). While the CH$_3$OH lines from
both torsionally ground and excited states are apparent toward the hot
core, only transitions from torsionally ground states are noticeable
at the latter position. Given that the torsionally excited states
require either high densities and temperatures for collisional
excitation, or strong infrared radiation fields for radiative pumping,
the above behavior readily demonstrates variations in excitation
conditions at different locations.

Figure \ref{trot} presents the derived rotational temperature
structure toward the Orion-KL region. To increase the S/N ratio, we
have smoothed/binned the original $0.1''$ pixel images to a larger
pixel size of $0.9''$, of the order of the beam size. Still, this map
gives one of the most detailed spatial pictures of the temperature
distribution in this region. The derived rotational temperatures are
between 50 and 350\,K, with a mean value of $\approx$ 150\,K over the
region. Formal statistical errors from the fitting mostly range
between 10 and 50\,K. The temperatures we obtained are consistent with
previous studies based on (sub)mm molecular line emission (e.g.,
\citealt{blake1987,wilner1994,schilke1997b,wilson2000,comito2005}) and
mid-infrared observations \citep{gezari1998}.  The highest temperature
locations are mostly found to be close to the hot core positions such
as the main submm continuum peak. In some regions, such as toward the
south-west of SMA1 near the compact ridge, lines from the torsionally
ground and excited states have similar intensities and thus appear to
be optically thick (Fig.~\ref{sample}). In such optically thick cases,
the fitting becomes insensitive to the rotational temperature along
the particular line of sight. Toward locations where more than half of
the CH$_3$OH lines show fitted opacities $>1$, we refrained from
temperature estimates and blanked the pixels in Figure
\ref{trot}. However, in the optical thick case the apparent intensity 
(brightness temperature) is directly related to true brightness
temperature, differing only by the filling factor. For example, using
the averaged temperature of 150 K and typical observed peak intensity
of 4\,Jy\,beam$^{-1}$ toward such optically thick directions, we find
filling factors of 0.3 or less in the smoothed data cube. For
positions where the spectral fitting is successful, the filling
factors are even smaller, 0.14 on average.  Such small filling factors
may appear somewhat surprising, given past studies. \citet{comito2005}
for example, adopted a source size of 10$''$ for CH$_3$OH emission but
also obtained a similar rotational temperature.  One expects that,
when imaged at a higher angular resolution, the CH$_3$OH emission
would be resolved and thus fill the smaller beam.  The fact that we
find rotational temperatures comparable to past studies but with
filling factor less than unity even for the optically thick lines
suggests that a significant amount of CH$_3$OH emission is missing and
the observed hot CH$_3$OH gas is actually clumpy. Indeed, smoothing
the data to the $20''$ CSO beam and comparing these spectra with the
single-dish observations by \citet{schilke1997b}, as much as 90$\%$ of
the integrated CH$_3$OH flux is resolved out by the interferometer.
Since the spatial filtering affects various CH$_3$OH lines
differently, this has to be taken into account as well. As discussed
in \S \ref{obs} \& \S \ref{sio}, the missing flux has to be
distributed over large-spatial scales, and the more interesting
parameter is the missing flux per synthesized beam.  Estimating the
latter values, we find that the CH$_3$OH missing flux per synthesized
SMA beam is in the worst case at a 10-20\% level, within the
calibration uncertainty. Therefore, it should not affect the
temperature determination significantly.

We note that there are a few caveats in our synthetic spectral fitting
process which does not always result in satisfactory fitted spectra.
As mentioned above, the fitting process fails toward optically thick
regions. Moreover, at some positions the CH$_3$OH lines appear to be
double-peaked, or a velocity offset between the torsionally ground
state and excited transitions is noticeable (Fig.~\ref{sample}). If
various kinematic components are present our fitting process often
fails to converge. Additionally, (minor) disagreement between the
observed and fitted spectra is not limited to the above cases but also
present in the two top spectra of Fig.~\ref{sample}. Since all the
transitions were observed simultaneously, calibration errors due to
poor pointing or amplitude scaling are likely to be canceled out.
Such deviations therefore suggest that additional non-LTE effects are
at work. Furthermore, radiative pumping of the torsionally excited as
well as the ground state transitions also needs to be considered,
although collisional excitation is likely to dominate in high density
regions like the Orion KL \citep{menten1986}.  A full statistical
equilibrium analysis such as suggested by \citet{leurini2004}, which
is outside the scope of this paper, and the inclusion of torsionally
excited states would be an interesting next step to investigate these
effects.

\subsection{Vibrational excited emission}

{\it SMA1:} The three vibrationally/torsionally excited lines in the
survey (CH$_3$OH, SO$_2$, HC$_3$CN) all show strong emission toward
the submm continuum source SMA1 (Fig.~\ref{images}). Since these lines
are usually excited by infrared radiation, this emission indicates a
deeply embedded infrared source at the position of SMA1. Similar
conclusions were drawn by \citet{devicente2002} from lower spatial
resolution Plateau de Bure observations of vibrational excited
HC$_3$N(10--9). Just based on the submm continuum study, we could not
judge well whether SMA1 is an independent protostellar source, or
whether it is an extension of the hot core
\citep{beuther2004g}. These vibrationally excited line studies 
support the independent protostellar object interpretation for
SMA1. Because SMA1 is detected neither at infrared nor at cm
wavelength, it is likely one of the youngest sources of the evolving
cluster.

{\it The compact ridge:} The torsionally excited CH$_3$OH is also
observed toward the compact ridge. We find only weak vibrationally
excited SO$_2$ and no HC$_3$N emission in this region. It is possible
that an additional embedded infrared source exists in the region of
the compact ridge, but we do not detect any emission above the noise
in the 865\,$\mu$m continuum there \citep{beuther2004g}. In contrast
to this, \citet{blake1996} detected weak mm continuum emission in that
region. However, there mm continuum map does not show a strong peak
but is more broadly distributed. The compact ridge is believed
to be the interface region between one (or more?) of the molecular
outflows with the dense ambient gas (e.g.,
\citealt{blake1987,liu2002}). Sine torsionally excited CH$_3$OH can also 
be produced in regions of extremely high densities (critical densities
of the order $10^{10}$\,cm$^{-3}$), it might also be possible that the
CH$_3$OH $v_t=1,2$ lines are excited in this interface region without
an embedded infrared source.

\section{Conclusion}

The presented observations comprise a large set of molecular line data
($\sim 145$ lines from $\sim 24$ species, isotopologues and
vibrational excited states) taken simultaneously during only two
observing nights. The large number of observed CH$_3$OH lines (46)
with $v_t$=0,1,2 are a powerful tool to derive physical parameters
(e.g., temperatures) of the molecular gas, and we find temperatures
between 50 and 350\,K throughout the region.

The SiO lines trace the collimated low-velocity molecular jet
emanating from source {\it I} as well as larger-scale emission likely
associated with a different outflow. In contrast to previously
reported ground state $^{28}$SiO(1--0) \& (2--1) maser emission close
to source {\it I}, we do not find detectable $^{28}$SiO(8--7) maser
emission there. Since source {\it n} is weak in the outflow-tracing
SiO emission, this could indicate that it may currently drive no
molecular outflow anymore.

Comparing different molecular species, for example, nitrogen-, oxygen-
and sulphur-bearing molecules, we find strong chemical gradients over
the observed region. For example, source {\it I} exhibits mainly SiO
emission whereas the hot core is especially strong in nitrogen-bearing
molecules like CH$_3$CN. The strongest emission features of most
oxygen-bearing molecules are south of the hot core toward the
so-called compact ridge. These differences have important implications
for studies lower-spatial-resolution studies and/or of sources at
larger distances. For example, analyzing data of a typical massive
star-forming region at approximately 4\,kpc distance and deriving
spatially averaged temperatures from the CH$_3$OH data does not
necessarily imply that the same temperatures can be attributed to the
(sub)mm dust continuum cores. Furthermore, the imaging of spectral
lines can be used as an additional tool for line
identifications. Since new interstellar molecular line identifications
get more complicated for large and complex molecules (see, e.g., the
controversy about glycine, \citealt{kuan2003,snyder2005}), imaging
each molecular line and comparing their spatial distributions is an
important consistency check for rigorous line identifications. 

All vibrationally/torsionally excited lines show strong emission
toward the submm continuum peak SMA1. Because this emission is usually
excited by infrared emission, it supports the notion that the recently
identified source SMA1 is likely of protostellar nature.

This dataset is particularly rich and has potential for follow-up
investigations. To mention just a couple: Imaging and potentially
identifying the low-level emission peaks, or better constraining the
underlying physical processes, which are responsible for the observed
chemical gradients, is of great interest.

\acknowledgments{We like to thank all people at the SMA for their
large efforts over many years to get the instrument going! An
additional Thank You to the referee Peter Schilke for detailed
comments improving the paper.  H.B. acknowledges financial support by
the Emmy-Noether-Program of the Deutsche Forschungsgemeinschaft (DFG,
grant BE2578/1).}


\noindent {\it All figure captions are directly below the
figures. Just for Fig.~\ref{images}, where the caption did not
properly fit on the page, we set it here.}

\noindent Caption to Fig.~\ref{images}.--- Continuum-subtracted 
images of molecular species: the thick contours present the molecular
emission (labeled in each panel) from 10 to 90\% (step 10\%) of each
species' peak emission. Additional parameters for each line image are
given in Table \ref{image_parameters}. The axis are offsets in the
directions of Right Ascension and Declination. The dashed contours
show the negative emission from 10 to 90\% (step 20\%) which is due to
missing short spacings. Source {\it I}, the hot core (HC), SMA1,
source {\it n}, IRc6 and the compact ridge are marked in each panel
and labeled in the $^{30}$SiO panel (2nd at top). Two additional
CH$_3$CN positions, where spectra are shown from in
Fig.~\ref{spectra_positions}, are marked with two additional stars in
the CH$_3$CN panel. The top-two rows show the Si- and and S-bearing
species, the middle-two rows present the OH-bearing molecules, and the
bottom-two rows focus on the N-bearing species. Lines with tentative
or uncertain identifications are marked with ``?'' (see Tables
\ref{lines} \& \ref{image_parameters}).

\begin{figure}
\begin{center}
\end{center}
\caption{Sub-mm continuum image of the Orion-KL region at 865\,$\mu$m
\citep{beuther2004g}. The contouring starts at the $2\sigma$ level of
70\,mJy/beam and continues in $2\sigma$ steps. The white crosses mark
the radio positions of sources {\it I} and {\it n} (two positions for
{\it n} because of its bipolar nature, \citealt{menten1995}). The
black stars show the infrared positions of source {\it n} and IRc2A-D
\citep{dougados1993}. The synthesized beam is shown at the bottom left
($0.78''\times 0.65''$). We did not detect any submm continuum feature
outside this central region.}
\label{continuum}
\end{figure}

\begin{figure}
\begin{center}
\end{center}
\caption{The left panel shows the uv-coverage from the two observed 
tracks. The right panel then presents an image of the dirty beam. The
contour levels are -0.125 (dashed) and 0.125 to 1 in 0.125 level steps
(full lines).}
\label{dirty}
\end{figure}

\begin{figure}
\begin{center}
\end{center}
\caption{Vector-averaged upper and lower sideband spectra in the
uv-domain on a baseline of $\sim 21$\,m. UL marks unidentified lines
and ``?'' only tentatively identified lines.}
\label{spectra_all}
\end{figure}

\begin{figure}
\begin{center}
\end{center}
\caption{Vector-averaged lower sideband spectra (ZOOM) in the
uv-domain on a short baseline of $\sim 21$\,m. The spectral resolution
is 0.7\,km\,s$^{-1}$. UL marks unidentified lines and ``?'' only
tentatively identified lines.}
\label{spectra_lower}
\end{figure}

\begin{figure}
\begin{center}
\end{center}
\caption{Vector-averaged upper sideband spectra (ZOOM) in the
uv-domain on a short baseline of $\sim 21$\,m. The spectral resolution
is 0.7\,km\,s$^{-1}$. UL mark unidentified lines and ``?'' only
tentatively identified lines.}
\label{spectra_upper}
\end{figure}

\begin{figure}
\begin{center}
\end{center}
\caption{Caption before Fig.~1.}
\label{images}
\end{figure}

\begin{figure}
\begin{center}
\end{center}
\caption{Spectra taken toward selected positions after imaging the
whole data cube, the spectral resolution is smoothed to
3\,km\,s$^{-1}$. The spatial positions are marked in
Fig.~\ref{images}. The X-axis shows the frequency [GHz] and the Y-axis
the amplitude [Jy]. Negative features are not real but caused by
missing short spacings. The positive emission is affected by the
missing short spacings as well.}
\label{spectra_positions}
\end{figure}

\begin{figure}
\begin{center}
\end{center}
\caption{$^{28}$SiO(8--7) channel map in grey-scale (the wedge is in
units of Jy), positive features are presented in full and negative in
dashed contours (contour levels $\pm0.5$, $\pm1$, $\pm1.5$, $\pm2$,
$\pm2.5$, $\pm3$, $\pm4$, $\pm5$, 6, and 7\,Jy). The channel-center
velocities are shown at the top left of each panel, the four dotted
crosses mark the positions of source {\it I}, the hot core, SMA1, and
source {\it n} (see Fig.~\ref{continuum}). The synthesized beam is
presented at the bottom left of each panel.}
\label{sio_channel}
\end{figure}

\begin{figure}
\begin{center}
\end{center}
\caption{$^{30}$SiO(8--7) channel map in grey-scale (the wedge is in
units of Jy), positive features are presented in full and negative in
dashed contours (contour levels are 0.15, $\pm0.3$, $\pm0.6$,
$\pm0.9$, $\pm1.2$, 1.5, 2, 3, 4, 5\,Jy). The channel-center
velocities are shown at the top left of each panel, the four dotted
crosses mark the positions of source {\it I}, the hot core, SMA1, and
source {\it n} (see Fig.~\ref{continuum}).}
\label{30sio_channel}
\end{figure}

\begin{figure}
\begin{center}
\end{center}
\caption{Vector-averaged $^{28}$SiO(8--7) spectrum observed on a short
baseline of $\sim 21$\,m with a velocity resolution of
0.7\,km\,s$^{-1}$.}
\label{sio_spectrum}
\end{figure}

\begin{figure}
\begin{center}
\end{center}
\caption{$^{28}$SiO(8--7) images integrated over selected velocity
ranges. The left panel presents the $^{28}$SiO emission at the central
velocities between 5 and 10\,km\,s$^{-1}$. The middle panel shows the
main $^{28}$SiO outflow, blue (solid contours) [-5,5]\,km\,s$^{-1}$ \&
red (dashed) [10,20]\,km\,s$^{-1}$. The right panel then shows the
$^{28}$SiO emission integrated over the spectral features at even
higher velocities (Fig.~\ref{sio_spectrum}), blue (solid)
[-18,-5]\,km\,s$^{-1}$ \& red (dashed) [23,35]\,km\,s$^{-1}$. Again,
to avoid confusion, negative emission due to missing short spacings is
omitted in these plots. Source {\it I}, the hot core (HC), SMA1,
source {\it n}, IRc6 and the compact ridge are marked in each panel
and labeled in the left panel (see Fig.~\ref{images}). The synthesized
beam is presented at the bottom left of each panel.}
\label{sio_sample}
\end{figure}

\begin{figure}
\begin{center}
\end{center}
\caption{$^{28}$SiO(8--7) (full line) and $^{30}$SiO(8--7) (dashed
line) spectra integrated over the central $4''\times 4''$ area.}
\label{sio_spectra}
\end{figure}

\begin{figure}
\begin{center}
\end{center}
\caption{CH$_3$OH spectra toward selected positions in
Orion-KL. The two top panels show observed spectra in black lines
(baselines at intensity 0) and corresponding fits in grey lines
(baselines at intensity 2.5) toward the Hot Core and the southern
Compact Ridge position marked in Figure \ref{trot}. The two
bottom panels present spectra that were not possible to fit
because they either show double peaks or are optically thick, the
positions are also marked in Fig.~\ref{trot}.}
\label{sample}
\end{figure}

\begin{figure}
\begin{center}
\end{center}
\caption{The rotational temperatures derived from the whole
lower-sideband CH$_3$OH data are presented in grey-scale. The contour
overlay shows the same CH$_3$OH ground state map as shown in Figure
\ref{images}. The triangles mark the positions of source {\it I}, the
mm continuum peak of the hot core, the continuum peak SMA1 and source
{\it n} (compare Fig.~\ref{continuum}). The squares mark additional
positions used to extract the spectra shown in Figure \ref{sample}
(CR: (Southern) Compact Ridge (see HCOOH map by \citealt{liu2002}),
OT: Optical thick, DP: Double peaked). To calculate the rotational
temperature map, we binned the original $0.1''$ pixel line maps to a
larger pixel size of $0.9''$ (to increase the S/N ratio). Therefore,
the temperature map gives values over a bit larger spatial area than
the not binned, higher-spatial-resolution CH$_3$OH line map.}
\label{trot}
\end{figure}

\clearpage

\begin{deluxetable}{lr}
\tablecaption{Observed lines \label{lines}}
\tablewidth{0pt}
\tablehead{
\colhead{$\nu$} & \colhead{line} \\
\colhead{GHz} & \colhead{} 
}
\startdata
337.252 & CH$_3$OH$(7_{5,2}-6_{5,2})$A($v_t$=2)\\
337.274 & CH$_3$OH$(7_{6,2}-6_{5,1})$A($v_t$=2)\\
337.279 & CH$_3$OH$(7_{3,5}-6_{2,4})$E($v_t$=2)\\
337.284 & CH$_3$OH$(7_{4,4}-6_{3,3})$A($v_t$=2)\\
337.297 & CH$_3$OH$(7_{4,3}-6_{4,3})$A($v_t$=1)\\
337.312 & CH$_3$OH$(7_{3,4}-6_{3,4})$A($v_t$=2)\\
337.348 & CH$_3$CH$_2$CN$(38_{3,36}-37_{3,35})$\\
337.397 & C$^{34}$S(7--6)\\
337.421 & CH$_3$OCH$_3(21_{2,19}-20_{3,18})$\\
337.446 & CH$_3$CH$_2$CN$(37_{4,33}-36_{4,32})$\\
337.464 & CH$_3$OH$(7_{7,1}-6_{6,0})$A\\
337.473 & UL\\
337.490 & HCOOCH$_3(27_{8,20}-26_{8,19})$E \&\\
        & CH$_3$OH$(7_{1,7}-6_{0,6})$E\\
337.503 & HCOOCH$_3(27_{8,20}-26_{8,19})$A\\
337.519 & CH$_3$OH$(7_{5,2}-6_{5,2})$E($v_t$=1)\\
337.546 & CH$_3$OH$(7_{6,1}-6_{6,1})$A($v_t$=1)\\
337.580 & $^{34}$SO$(8_8-7_7)$\\
337.605 & CH$_3$OH$(7_{3,5}-6_{2,4})$E($v_t$=1)\\
337.611 & CH$_3$OH$(7_{2,5}-6_{2,5})$E($v_t$=1)\\
337.626 & CH$_3$OH$(7_{5,3}-6_{4,2})$A$^+$($v_t$=1)\\
337.636 & CH$_3$OH$(7_{5,3}-6_{4,2})$A$^-$($v_t$=1)\\
337.642 & CH$_3$OH$(7_{4,3}-6_{4,3})$A($v_t$=1)\\
337.644 & CH$_3$OH$(7_{4,4}-6_{3,3})$E($v_t$=1)\\
337.646 & CH$_3$OH$(7_{2,6}-6_{1,5})$A($v_t$=1)\\
337.648 & CH$_3$OH$(7_{1,6}-6_{1,6})$A($v_t$=1)\\
337.655 & CH$_3$OH$(7_{5,2}-6_{5,2})$A($v_t$=1)\\
337.671 & CH$_3$OH$(7_{5,3}-6_{4,2})$A($v_t$=1)\\
337.685 & CH$_3$OH$(7_{6,1}-6_{6,1})$A($v_t$=1)\\
337.708 & CH$_3$OH$(7_{3,4}-6_{3,4})$A($v_t$=1)\\
337.712 & CH$_3$OCH$_3(7_{4,4}-6_{3,3})$EA\\
337.722 & CH$_3$OCH$_3(7_{4,4}-6_{3,3})$EE\\
337.732 & CH$_3$OCH$_3(7_{4,3}-6_{3,3})$EE\\
337.749 & CH$_3$OH$(7_{4,4}-6_{3,3})$A($v_t$=1)\\
337.771 & CH$_3$OCH$_3(7_{4,4}-6_{3,4})$EA\\
337.778 & CH$_3$OCH$_3(7_{4,4}-6_{3,4})$EE\\
337.787 & CH$_3$OCH$_3(7_{4,3}-6_{3,4})$AA\\
337.830 & HC$_3$N(37--36)($v_7=1$)@337.825?\\
337.838 & CH$_3$OH$(20_{6,14}-21_{5,15})$ \\
337.878 & CH$_3$OH$(7_{4,3}-6_{4,3})$A($v_t$=2)\\
337.892 & SO$_2$$(21_{2,20}-21_{1,21})v_2=1$\\
337.915 & UL\\
337.939 & UL\\
337.969 & CH$_3$OH$(7_{4,3}-6_{4,3})$A($v_t$=1)\\
338.081 & H$_2$CS$(10_{1,10}-9_{1,9})$\\
338.106 & UL\\
338.125 & CH$_3$OH$(7_{4,4}-6_{3,3})$E\\
338.144 & CH$_3$CH$_2$CN$(37_{3,34}-36_{3,33})$ \\
338.214 & CH$_2$CHCN$(37_{1,37}-36_{1,36})$\\
338.306 & SO$_2$$(14_{4,14}-18_{3,15})$, many comp.\\
338.345 & CH$_3$OH$(7_{4,0}-6_{3,4})$E\\
338.356 & HCOOCH$_3(27_{8,19}-26_{8,18})$A\\
338.396 & HCOOCH$_3(27_{7,21}-26_{7,20})$E\\
338.405 & CH$_3$OH$(7_{7,1}-6_{6,0})$E\\
338.409 & CH$_3$OH$(7_{4,4}-6_{3,3})$A\\
338.415 & part of CH$_3$OH$(7_{4,4}-6_{3,3})$A?\\
338.431 & CH$_3$OH$(7_{1,7}-6_{0,6})$E\\
338.442 & CH$_3$OH$(7_{7,1}-6_{0,0})$A\\
338.448 & CH$_2$CHCN$(37_{0,37}-36_{0,36})$\\
338.457 & CH$_3$OH$(7_{1,6}-6_{1,6})$E\\
338.461 & part of CH$_3$OH$(7_{1,6}-6_{1,6})$E?\\
338.475 & CH$_3$OH$(7_{6,1}-6_{6,1})$E\\
338.486 & CH$_3$OH$(7_{6,1}-6_{6,1})$A\\
338.491 & part of CH$_3$OH$(7_{6,1}-6_{6,1})$A?\\
338.504 & CH$_3$OH$(7_{2,6}-6_{1,5})$E\\
338.510 & part of CH$_3$OH$(7_{2,6}-6_{1,5})$E?\\
338.513 & CH$_3$OH$(7_{6,2}-6_{5,1})$A\\
338.517 & SO$_2$$(55_{6,50}-54_{7,47})\,v_2=1$  \\
338.530 & CH$_3$OH$(7_{6,2}-6_{5,1})$E\\
338.541 & CH$_3$OH$(7_{5,2}-6_{5,2})$A$^+$\\
338.543 & CH$_3$OH$(7_{5,2}-6_{5,2})$A$^-$\\
338.547 & part of CH$_3$OH$(7_{5,2}-6_{5,2})$A$^-$?\\
338.560 & CH$_3$OH$(7_{2,5}-6_{2,5})$E   \\
338.565 & part of CH$_3$OH$(7_{2,5}-6_{2,5})$E?\\
338.583 & CH$_3$OH$(7_{5,2}-6_{5,2})$E\\
338.587 & part of SO$_2$/CH$_3$OH?\\
338.592 & part of SO$_2$/CH$_3$OH?\\
338.607 & part of SO$_2$/CH$_3$OH?\\
338.612 & SO$_2$$(20_{1,19}-19_{2,18})$\\
338.615 & CH$_3$OH$(7_{4,3}-6_{4,3})$E\\
338.621 & part of CH$_3$OH$(7_{4,3}-6_{4,3})$E?\\
338.640 & CH$_3$OH$(7_{5,3}-6_{4,2})$A\\
338.646 & part of CH$_3$OH$(7_{5,3}-6_{4,2})$A?\\
338.723 & CH$_3$OH$(7_{3,5}-6_{2,4})$E\\
338.729 & part of CH$_3$OH$(7_{3,5}-6_{2,4})$E?\\
338.760 & $^{13}$CH$_3$OH$(13_{7,7}-12_{7,6})$A\\
338.772 & HC$_3$N$(37-36)$($v_7=2$)@338.768?\\
338.782 & part of $^{34}$SO$_2(14_{4,10}-14_{3,1})$?\\
338.786 & $^{34}$SO$_2(14_{4,10}-14_{3,1})$\\                           
338.884 & C$_2$H$_5$OH$(15_{7,9}-15_{6,10})$?$^b$\\
338.930 & $^{30}$SiO(8--7), broad line\\
339.058 & C$_2$H$_5$OH$(14_{7,7}-14_{6,8})$?$^b$\\
339.071 & UL\\
339.129 & HCOOCH$_3(13_{7,7}-12_{6,7})$E\\
339.138 & $^{13}$CH$_3$CN$(19_{6,19}-18_{6,18})$@339.137 \\
        & \& C$_2$H$_5$OH$(27_{7,20}-27_{3,24})$@339.141?$^b$\\
339.153 & HCOOCH$_3(13_{7,6}-12_{6,6})$E\\
347.331 &  $^{28}$SiO(8--7), broad line \\
347.438 &  UL \\
347.446 &  UL \\
347.478 &  HCOOCH$_3(27_{1,26}-26_{1,25})$E \\
347.494 &  HCOOCH$_3(27_{5,22}-26_{5,21})$A \\
347.516 &  UL \\
347.570 & UL \\
347.590 &  HCOOCH$_3(16_{6,10}-15_{5,11})$A\\
347.599 &  HCOOCH$_3(16_{6,10}-15_{5,11})$E\\
347.605 &  HCOOCH$_3(28_{10,18}-27_{10,17})$E\\
347.617 &  HCOOCH$_3(28_{10,19}-27_{10,18})$A\\
347.628 &  HCOOCH$_3(28_{10,19}-27_{10,18})$E\\
347.667 &  UL \\
347.699 &  CH$_3$CH$_2$$^{13}$CN$(9_{8,1}-9_{7,2})$?\\
347.759 &  CH$_2$CHCN$(36_{2,34}-35_{2,32})$\\
347.846 &  $^{13}$CH$_3$OH$(21_{12,9}-20_{12,8})$A\\
347.903 &  UL \\
347.916 &  C$_2$H$_5$OH$(20_{4,17}-19_{4,16})$?$^b$\\
347.928 &  UL \\
348.050 &  HCOOCH$_3(28_{4,24}-27_{4,23})$E\\
348.066 &  HCOOCH$_3(28_{6,23}-27_{6,22})$A\\
348.084 & UL, CH$_2$CHCN$(57_{2,55}-58_{0,58})$? \\
348.101 &  $^{13}$CH$_3$OH$(11_{6,6}-10_{6,5})$E\\
348.118 &  $^{34}$SO$_2(19_{4,16}-19_{3,17})$\\
348.161 & UL, CH$_2$CHCN$(38_{0,38}-37_{0,37})$? \\
348.200 &  CH$_2$CHCN$(25_{7,19}-26_{6,20})$?\\
348.261 &  CH$_3$CH$_2$CN$(39_{2,37}-38_{2,36})$ \\
348.340 &  HN$^{13}$C(4--3), blended \\
348.345 &  CH$_3$CH$_2$CN$(40_{2,39}-39_{2,38})$, blended\\
348.388 &  SO$_2(24_{2,22}-23_{3,21})$ \\
348.519 &  UL$^a$, HNOS$(1_{1,1}-2_{0,2})$@348.518?\\
348.532 &  H$_2$CS$(10_{1,9}-9_{1,8})$\\
348.553 &  CH$_3$CH$_2$CN$(40_{1,39}-39_{1,38})$ \\
348.647 &  CH$_2$CHCN$(20_{4,17}-21_{2,20})$? \\
348.692 &  UL \\
348.723 &  HCOOCH$_3$$(28_{9,20}-27_{9,18})$E?\\
348.785 &  CH$_3$CN$(19_{10}-18_{10})$ \\
348.847 &  C$_2$H$_5$OH$(10_{6,4}-9_{5,4})$?$^b$\\
348.910 &  HCOOCH$_3(28_{9,20}-27_{9,19})$E\\
348.911 & CH$_3$CN$(19_{9}-18_{9})$\\
348.915 & HCOOCH$_3(28_{9,20}-27_{9,19})$A\\
348.966 & UL \\
348.991 & CH$_2$CHCN$(37_{1,36}-36_{1,35})$ \\
349.025 & CH$_3$CN$(19_{8}-18_{8})$  \\
349.049 & HCOOCH$_3(28_{9,19}-27_{9,18})$E\\
349.066 & HCOOCH$_3(28_{9,19}-27_{9,17})$A  \\
349.107 & CH$_3$OH$(14_{8,7}-14_{7,7})$\\
349.114 & CH$_3$$^{13}$CN$(19_{5}-18_{5})$?\\
349.125 & CH$_3$CN$(19_{7}-18_{7})$ \\
349.173 & CH$_3$$^{13}$CN$(19_{4}-18_{4})$?\\
\enddata
\tablenotetext{a}{\footnotesize The spatial
distributions resembles nitrogen-bearing molecules like CH$_3$CN.}
\tablenotetext{b}{\footnotesize The spatial distributions of the
C$_2$H$_5$OH emission are not the same for all lines. Thus, some are
probably misassigned, it is difficult to judge which.}
\tablecomments{\footnotesize{Lines marked with ``?'' are only
tentatively identified. UL: Unidentified line}}
\end{deluxetable}

\begin{deluxetable}{rrr}
\tablecaption{Detected molecular species \label{species}}
\tablewidth{0pt}
\tablehead{
\colhead{Species} & \colhead{Isotopologues} & \colhead{Vib. states} 
}
\startdata
$^{28}$SiO & $^{30}$SiO \\
C$^{34}$S\\
$^{34}$SO \\
SO$_2$ & $^{34}$SO$_2$ & SO$_2$ $v_2=1$ \\
HN$^{13}$C \\
H$_2$CS\\
  & & HC$_3$N $v_7=1,2$ ? \\
CH$_3$OH & $^{13}$CH$_3$OH & CH$_3$OH $v_t=1,2$ \\
CH$_3$CN & $^{13}$CH$_3$CN \\
         & CH$_3^{13}$CN? \\
CH$_2$CHCN \\
HCOOCH$_3$ \\
CH$_3$OCH$_3$ \\
C$_2$H$_5$OH \\
CH$_3$CH$_2$CN & CH$_3$CH$_2$$^{13}$CN? \\
\enddata
\tablecomments{\footnotesize Species marked with ``?'' are only tentatively identified.}
\end{deluxetable}

\begin{deluxetable}{lrrr}
\tablecaption{Line image parameters to Fig.~\ref{images} \label{image_parameters}}
\tablewidth{0pt}
\tablehead{
\colhead{Line} & \colhead{$E_{\rm{upper}}$} & \colhead{$v_{\rm{low}},v_{\rm{high}}$} & \colhead{$S_{\rm{peak}}$} \\
\colhead{} & \colhead{[K]} & \colhead{[km\,s$^{-1}$]} & \colhead{[Jy]} 
}
\startdata
$^{28}$SiO(8--7) & 75 & -20,40 & 2.1 \\
$^{30}$SiO(8--7) & 73 & -12,28 & 4.0 \\
C$^{34}$S(7--6) & 65 & -2,14 & 2.4 \\
H$_2$CS$(10_{1,10}-9_{1,9})$ & 105 & 2,10 & 3.4 \\
SO$_2(24_{2,22}-23_{3,21})$ & 293 & -10,20 & 2.3 \\
SO$_2(21_{2,20}-21_{1,21})(v_2=1)$ & 965 & 4,10$^a$ & 0.6 \\
$^{34}$SO$_2(19_{4,16}-19_{3,17})$ & 213 & 3,13 & 1.6 \\
$^{34}$SO$(8_8-7_7)$ & 86 & 1,15 & 3.3 \\
CH$_3$OH$(7_{5,2}-6_{5,2})$E & 112 & 2,6$^a$ & 3.3 \\
$^{13}$CH$_3$OH$(13_{7,7}-12_{7,6})$A & 206 & 4,12 & 2.0 \\
CH$_3$OH$(7_{4,3}-6_{3,3})$A($v_t=1$) & 390 & 3,13 & 2.8 \\
CH$_3$OH$(7_{5,2}-6_{5,2})$A($v_t=2$) & 722 & 4,12 & 1.6 \\
HCOOCH$_3(28_{10,19}-27_{10,18})$ & 307 & 4,10 & 2.6 \\
CH$_3$OCH$_3(7_{4,3}-6_{3,3})$EE & 48 & 5,10 & 3.3 \\
HCOOH$(15_{4,12}-14_{4,11})^b$ or & 181 & 0,8 & 3.0 \\
CH$_3$CH$_2$CN$(37_{3,34}-36_{3,33})$\\
? C$_2$H$_5$OH$(15_{7,9}-15_{6,10})$ & 162 & 3,7 & 1.1 \\
CH$_3$CN$(19_8-18_8)$ & 624 & 1,9 & 2.0 \\
$^{13}$CH$_3$CN$(19_{6,19}-18_{6,18})$ & 420 & 4,10$^a$ & 0.9 \\
? CH$_3^{13}$CN$(19_5-18_5)$ & 346 & 2,10 & 5.1 \\
CH$_2$CHCN$(37_{1,37}-36_{1,36})$ & 312 & 2,10 & 1.7 \\
CH$_3$CH$_2$CN$(39_{2,37}-38_{2,36})$ & 344 & -4,12 & 1.3 \\
? CH$_3$CH$_2^{13}$CN$(9_{8,1}-9_{7,2})$ & 90 & 6,12 & 0.9 \\
? HC$_3$N(37--36)($v_7=1$) & 628 & 1,11 & 2.4 \\
? CH$_2$CHCN$(25_{7,19}-26_{6,20})$ & 746 & 6,11 & 1.0  
\enddata
\tablenotetext{a}{\footnotesize Small integration ranges to avoid line blending.}
\tablenotetext{b}{\footnotesize See \S\ref{results} and Fig.~\ref{images} for a discussion.}  
\tablecomments{\footnotesize Species marked with ``?'' are only tentatively identified.}
\end{deluxetable}

\begin{deluxetable}{lcccc}
\tablecaption{$^{28}$SiO and $^{30}$SiO line intensities \label{sio_intensities}}
\tablewidth{0pt}
\tablehead{
\colhead{Line} & \colhead{CSO$^{a}$($20''$)} & \colhead{SMA$^{b}$($20''$)} & \colhead{SMA$_{>0}^{c}$($20''$)} & \colhead{Range} \\
\colhead{} & \colhead{[K]} & \colhead{[K]} & \colhead{[K]} & \colhead{[\%]} 
}
\startdata
$^{28}$SiO & 34.5 & 0.1 & 4.4 & 0.3-13 \\
$^{30}$SiO & 4.1  & 0.6 & 2.0 & 15-49
\enddata
\tablenotetext{a}{\footnotesize From \citet{schilke1997b}.}
\tablenotetext{b}{\footnotesize Convolved to the $20''$ CSO beam.}
\tablenotetext{c}{\footnotesize Convolved to the $20''$ CSO beam after
masking out the negative emission.}
\end{deluxetable}

\end{document}